\documentclass[twocolumn]{aastex631}

\usepackage{natbib}

\usepackage{graphicx}
\usepackage{amssymb}
\usepackage{multirow}
\usepackage{float}

\defcitealias{Ho2021}{H+21} 	
\defcitealias{HeinkeHo2010}{HH10} 	
\defcitealias{Ho2009}{HH09} 	
\defcitealias{Elshamouty2013}{E+13}
\defcitealias{Posselt2013}{P+13}
\defcitealias{Posselt2018}{PP18}

\newcommand{\be}{\begin{equation}}
\newcommand{\ee}{\end{equation}}

\shorttitle{The cooling of the Cas A CCO from 2006 to 2020}
\shortauthors{Posselt et al.}

\begin{document}
 
\title{The cooling of the Central Compact Object in Cas A from 2006 to 2020}
\author{B. Posselt}
\correspondingauthor{B.Posselt}
\affil{Department of Astrophysics, University of Oxford, Denys Wilkinson Building, Keble Road, Oxford OX1 3RH, UK}
\affil{Department of Astronomy \& Astrophysics, Pennsylvania State University, 525 Davey Lab,University Park, PA 16802, USA}
\email{posselt@psu.edu}

\author{G. G. Pavlov}
\affil{Department of Astronomy \& Astrophysics, Pennsylvania State University, 525 Davey Lab,University Park, PA 16802, USA}

\begin{abstract}
We report on the study of six \emph{Chandra} observations (four epochs) of the Central Compact Object (CCO) in the Cassiopeia A supernova remnant with the ACIS instrument in the subarray mode. This mode minimizes spectrum-distorting instrumental effects such as pileup. The data were taken over a time span of $\sim 14$ years. 
If a non-magnetic carbon atmosphere is assumed for this youngest known CCO, then the temperature change is constrained to be $\dot{T}=-2900\pm 600$\,K\,yr$^{-1}$ or $\dot{T}=-4500\pm 800 $\,K\,yr$^{-1}$ ($1\sigma$ uncertainties) for constant or varying absorbing hydrogen column density. These values correspond to cooling rates of $-1.5 \pm 0.3$\,\% per 10\,yr and  $-2.3 \pm 0.4$\,\% per 10\,yr, respectively.
We discuss an apparent increase in the cooling rate in the last five years and the variations of the inferred absorbing hydrogen column densities between epochs. 
Considered together, these changes could indicate systematic effects such as caused by, e.g., an imperfect calibration of the increasing contamination of the ACIS filter.
\end{abstract}


\section{Introduction}
\label{intro}
The Central Compact Object (CCO) in the Cassiopeia A (Cas\,A) supernova remnant is the youngest known ($\approx 300$\,yr) neutron star in our Galaxy with an apparently purely thermal X-ray spectrum. The thermal evolution of such a young neutron star is interesting because the cooling rate at this age strongly depends on poorly known properties in the neutron star interior, in particular superfluidity
(e.g., \citealt{Shternin2021, Page2004, Yakovlev2004}). 
Over the course of their evolution, CCOs are discussed to exhibit changes in their atmosphere compositions -- e.g., \citet{Chang2004}; \citet[H+21 in the following]{Ho2021}. According to one model, they have buried, and then reemerging magnetic fields that are associated with increasing temperatures (e.g., \citealt{Ho2011}). As the youngest member of this class, the Cas\,A CCO is an important reference point for such studies.\\

The cooling of this CCO has been the topic of continued interest and X-ray monitoring after       
\citet{HeinkeHo2010} 
reported an unexpectedly rapid 4\% ($5.4\sigma$) decline of the surface temperature and a 21\% flux decline over the time span of 10 years. These results were based on spectral fits using a non-magnetic carbon atmosphere model that covers the whole neutron star surface and has a uniform effective temperature. The data were obtained from \emph{Chandra} observations with ACIS-S in the Graded mode, aimed primarily at the study of the supernova remnant.
Including more observation epochs in this mode and improved calibration data, \citet{Elshamouty2013, Wijngaarden2019} and \citetalias{Ho2021} presented updated cooling rates. 
Based on 19 years of Graded mode ACIS-S data, the most recent ten-year cooling rates are $-2.2 \pm 0.2 (1\sigma)$\,\% and $-2.8 \pm 0.3 (1\sigma)$\,\% for constant and varying $N_H$, the  absorbing hydrogen column density, respectively if a carbon atmosphere model is assumed \citepalias{Ho2021}.\\

These \emph{Chandra} ACIS-S Graded mode observations of the CCO suffered from several instrumental effects. 
Since the primary target was the supernova remnant, these observations used the full ACIS-S3 chip {which led to relatively slow readout}.
Since Cas\,A is bright, the slow readout implies that photon pileup is the most important instrumental effect. Pileup means that two or more photons are detected as a single event\footnote{For more details, see \url{cxc.harvard.edu/ciao/ahelp/acis_pileup.html}}. Photon pileup distorts the observed CCO spectrum. 
The pileup fraction of the Cas\,A data is gradually decreasing because the sensitivity of the ACIS detector decreases over time. This is mostly due to a contaminant accumulating on the optical-blocking filters of the ACIS detectors. In addition, not all X-ray events are telemetered in the Graded mode\footnote{For more details, see \url{cxc.harvard.edu/ciao/why/cti.html}}, which can also affect the spectrum.\\

The ACIS-S subarray mode avoids the spectral distortion effects due to photon pileup and the Graded telemetry mode. However, it cannot avoid the effect of the changing sensitivity of the ACIS detector due to the contamination.
Using this more suitable instrument mode, \citet{PavlovL2009}, \citet[P+13 in the following]{Posselt2013}, and \citet[PP18 in the following]{Posselt2018}
also carried out monitoring studies of the temperature evolution of the Cas\,A CCO. \citetalias{Posselt2018} reported conservative $3\sigma$ upper limits of $<3.3$\,\% and $<2.4$\,\% for the {absolute value of the} ten-year cooling rate if a non-magnetic carbon atmosphere model is assumed for varying or constant $N_H$, respectively.

\section{Observations and Data Reduction}
\label{obsred}
\begin{deluxetable}{lrllllrrr}[]
\vspace{0.1cm}
\tablecaption{Observation Parameters with CALDB 4.9.5\label{table:obs}}
\tablewidth{0pt}
\tablehead{
\colhead{ID} & \colhead{ObsID} & \colhead{MJD} & \colhead{$T_{\rm exp}$} & \colhead{$C$} & \colhead{$f_{\rm Src}$} & \colhead{S3$_X$} & \colhead{S3$_Y$} &  \colhead{$\theta$}\\
\colhead{} & \colhead{} & \colhead{} & \colhead{ks} & \colhead{cts} & \colhead{\%} & \colhead{pix} & \colhead{pix} & \colhead{$\arcsec$}\\
}
\startdata
P1  &  6690 & 54027 & 61.7 & 7443 & 86.5 & 210.7 & 49.0 & 18.4\\
P2  & 13783 & 56053 & 63.4 & 6773 & 87.3 & 215.2 & 50.7 & 17.2\\
P3a  & 16946 & 57140 & 68.1 & 6263 & 87.8 & 229.4 & 54.3 & 15.8\\
P3b  & 17639 & 57143 & 42.7 & 4556 & 82.5 & 574.6 & 508.1 & 174.2\\
P4a & 22426 & 58980 & 48.2 & 3859 & 86.7 & 334.2 & 506.0 & 57.2\\
P4b & 23248 & 58983 & 28.2 & 2107 & 85.6 & 334.1 & 506.5 & 57.1\\
\enddata
\tablecomments{The ID indicates the abbreviation used for the observing epoch of the \emph{Chandra} data set with the listed ObsID, $T_{\rm exp}$ is the dead-time-corrected exposure time after filtering {for high background}, (total) counts $C$ and the source count fraction $f_{\rm Src}$ correspond to the source extraction regions used for the spectral fits in Table~\ref{table:newcarbonfits}. S3$_X$ and S3$_Y$ are the centroid chip coordinates on ACIS-S3. $\theta$ is the off-axis angle.} 
\vspace{-0.8cm}
\end{deluxetable}

For this work, we use only \emph{Chandra} ACIS subarray mode observations. In the subarray mode, only a part of the ACIS chip is read out. 
The ACIS-S3 chip in the 100 pixel subarray is used for each observation listed in Table~\ref{table:obs}.
This subarray mode reduces the frame time to 0.34\,s versus the 3.24\,s in full-frame mode, reducing the pile-up fraction to less than 1.6\% in all epochs (compared to $\sim 20$\,\% in the case of the early full frame mode data, \citealt{PavlovL2009}).\\

Observing epochs P3a and P3b were obtained three days apart in May 2015, with the subarray placed near the chip readout in P3a, and at the center of the chip for P3b (see \citetalias{Posselt2018}  for more details; {\footnote{P3a and P3b were named P3 and P4 in \citetalias{Posselt2018} because they were independent programs with the CCO at different chip positions. Here, we emphasize their close proximity in time in comparison to the next observations.}}). In observing epochs P1 (2006) and P2 (2013), the subarray was also placed near the chip readout.  
The two new observations of P4 (2020), which we call P4a and P4b, were obtained 3 days apart in May 2020 with the subarray placed at the center of the chip.\\

We employ CIAO version 4.12 and the most recent CALDB version 4.9.5 for the data reduction and extraction of the spectra, and XSPEC (version 12.10.1) for the spectral analysis. The analysis is carried out in the same way as presented for epoch P1 to P3b  by \citetalias{Posselt2018} and \citetalias{Posselt2013}. In particular, we use similar extraction regions for the source and background. Intervening filaments of the supernova remnant are excluded from the background regions. Spectra are binned with a signal to noise ratio of at least 10 per energy bin.\\
 
In comparison to the previous CALDB versions used by \citetalias{Posselt2018}, the version 4.9.5 includes not only updates on the ACIS filter contamination correction,
but also on the aspect solution. This resulted in slightly changed off-axis angles (Table~\ref{table:obs}) for P1-P3b in comparison to \citetalias{Posselt2018}. 
Due to the changed contamination correction, the obtained spectral fit parameters are also slightly different as discussed in Section~\ref{sec:res}.
We note that the measured offsets between the CCO centroid positions in P1--P4 are still too small in comparison to the absolute astrometry uncertainty of \emph{Chandra}, $0\farcs{4}$.\\

For the spectral fits, we use the carbon atmosphere models by \citet{Suleimanov2014} with a negligibly low magnetic field ($B<10^{8}$\,G), a surface gravitational acceleration of $\log g=14.45$ and a gravitational redshift of $z=0.375$, which corresponds to a neutron star with $M_{\rm NS}=1.647$\,M$_{\odot}$ and  $R_{\rm NS}=10.33$\,km.
Using a distance of 3.4\,kpc ($d=3.4^{+0.3}_{-0.1}$\,kpc; \citealt{Reed1995}), 
we fix the
normalization,
$\mathcal{N}=R^2_{\rm NS} / d_{\rm 10kpc}^2=923$, where  $R_{\rm NS}$ is the neutron star radius in km, and $d_{\rm 10kpc}$ is the distance in 10\,kpc. 
{\citetalias{Posselt2013}} 
showed that the significance of the temperature (or flux) \emph{difference} is very similar to those obtained using tied normalizations (same emission size) or normalizations allowed to vary between observing epochs. The used spectral models are the same as in our previous works (\citetalias{Posselt2013}, \citetalias{Posselt2018}).
Here, we only use the non-magnetic carbon atmosphere model.
Hydrogen atmosphere models fit the data equally well, but require an emission area smaller than the total neutron star surface (see \citetalias{Posselt2013} for a detailed discussion).

\section{Results and Discussion}
\label{sec:res}
We verfied that the two epochs P3a and P3b, as well as the two epochs P4a and P4b give similar spectral fit results within uncertainties. 
This is not surprising
because each of this pair of observations is only 3 days apart and the CCO or its environment is unlikely to change over that time. We therefore tie the spectral fit parameters of P3a with those of P3b (epoch 3 in the following), and similarly P4a with P4b (epoch 4 in the following). 
As reference time in each epoch, we utilize the exposure-weighted average observing date.\\

\begin{figure}[]
{\includegraphics[width=86mm]{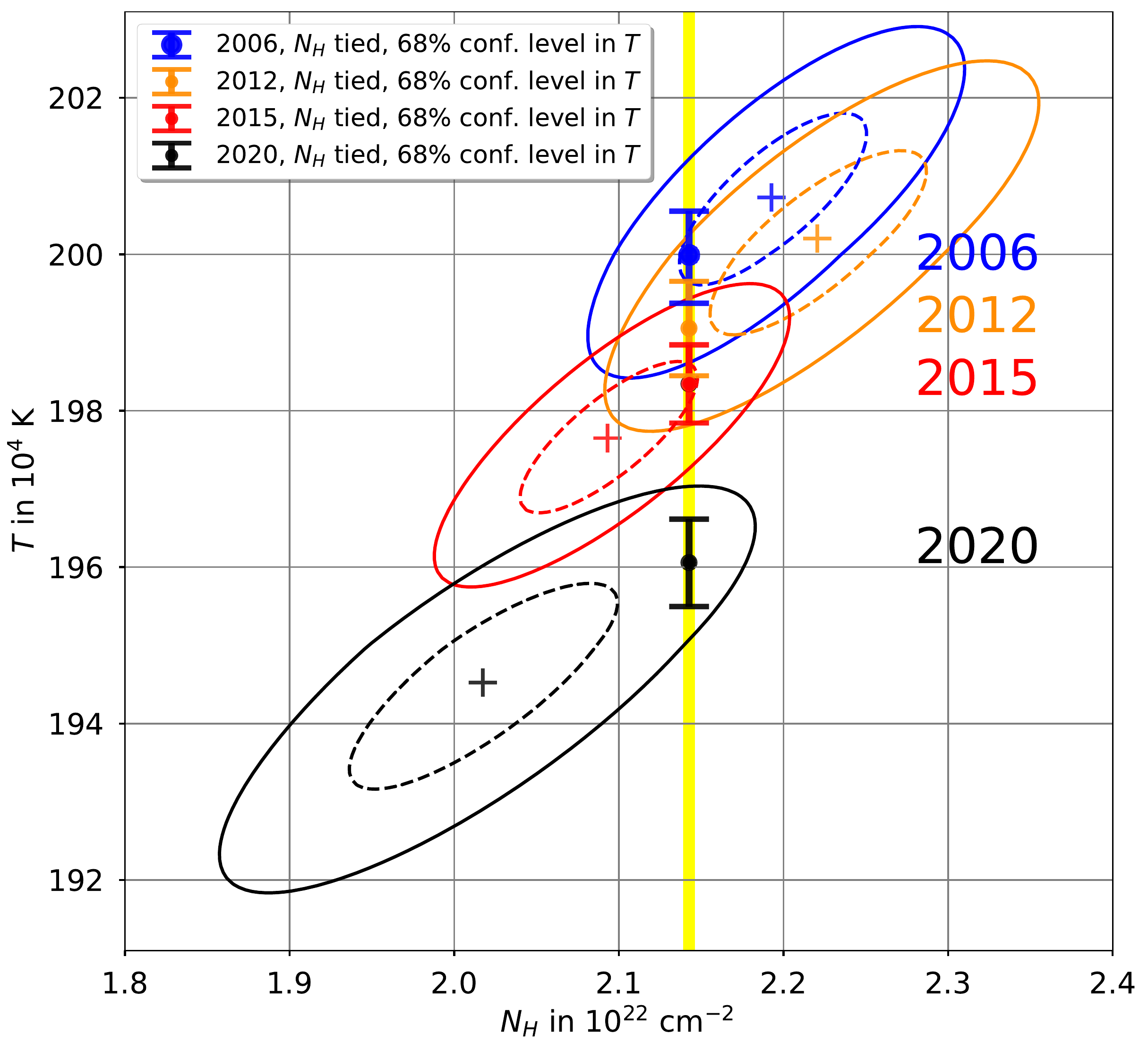}} 
\caption{Temperature versus $N_{\rm H}$ confidence contours (68\,\% -- dashed , 99\,\% -- solid) for the fit to the carbon atmosphere model with $\log g=14.45$, $z=0.375$. The best-fit values are marked with crosses. 
In the spectral model for this and the following figure, it is assumed that the whole neutron star surface is emitting in X-rays and that the distance is 3.4\,kpc (norms fixed). 
If the $N_{\rm H}$ is tied to the same value in all epochs (the respective fit result for $N_{\rm H}$ is marked with the yellow line), the obtained temperatures and their {68\% confidence level (i.e., the $1\sigma$ uncertainty for this one parameter of interest)} are plotted in the same colors as the confidence contours of the same data {for two parameters of interest}.
\label{carbonNHT}}
\end{figure}

Table~\ref{table:newcarbonfits} lists the reference times and best fitting parameters for the carbon atmosphere model. We consider two cases -- $N_H$ is tied to the same value for all epochs, or it is allowed to vary.
If $N_H$ is allowed to vary between the epochs, the results for each epoch are independent, and we can compare the results of the first three epochs with our earlier work to identify differences due to a changed calibration data base, i.e., CALDB version 4.9.5 in comparison to CALDB 4.7.3 utilized by \citetalias{Posselt2018}. 
The temperature values for each of the three epochs, obtained with the old and new CALDB, agree within $1\sigma$. The third epoch has the largest difference, with a lower temperature for CALDB 4.9.5.
The \emph{absolute} value of the temperature difference between epochs 1 and 2  (5.54 years) slightly decreases to $(0.5 \pm 1.7) \times 10^4$\,K (from the previous $(0.8 \pm 1.7) \times 10^4$\,K; 90\% confidence levels as in Table~\ref{table:newcarbonfits}) while between epochs 2 and 3 (only 2.98 years), the {absolute} temperature difference slightly increases to $(2.5 \pm 1.7) \times 10^4$\,K (from the previous $2.1^{+1.8}_{-1.7} \times 10^4$\,K; uncertainties are the 90\% confidence level). The best-fit absorbing hydrogen column densities $N_H$ also change slightly, with $N_H$ from epoch 2 and 
{4} being different by more than $3\sigma$.
This $N_H$ difference is also apparent in Figure~\ref{carbonNHT}.
If taken at face value, two interpretations are possible. One is that the hydrogen column density towards the CasA CCO decreased by 9\% in 8\,years. Then  the results of the fit for varying $N_H$ would be expected to be closer to reality than the fit result for tied $N_H$.
The other interpretation is that the effects of the ACIS filter contamination are not fully corrected in all epochs. 
This could introduce systematic errors which are partly counterbalanced by a slightly offset $N_H$ best-fit value. The general effects of the ACIS contamination correction on the fit values of the Cas A CCO have been established previously by introducing changes of the thickness in the contaminant model \citepalias{Posselt2013,Posselt2018}. However, detailed calibration anlaysis, carried out by \citet{Plucinsky2018}, showed more complicated behavior of the different contamination components with time, including their chemical composition, thickness, accumulation rate, and spatial distribution. An examination of the many contaminant parameters with respect to the influence on the CCOs fit parameters is beyond the scope of this report. However, we note that dedicated calbration observations, modeling and updates of the contaminant model are regularly carried out by the \emph{Chandra X-ray Center} calibration team, providing further improvements of the ACIS contamination model.

\begin{deluxetable*}{lllllllc}[t]
\tablecaption{Fit results for the carbon atmosphere models with $\log g=14.45$ and $z=0.375$ \label{table:newcarbonfits}}
\tablewidth{0pt}
\tablehead{
\colhead{Data} & \colhead{$t_{\rm Omid}$}  & \colhead{$N_{\rm H}$} & \colhead{$T_{\rm eff}$} &  \colhead{$F^{\rm{abs}}_{-13}$} & \colhead{$F^{\rm{unabs}}_{-12}$}  & \colhead{$L_{\rm bol}^{\infty}$} & \colhead{$\chi^2_{\nu}$ (dof)} \\
\colhead{ } & \colhead{yr } & \colhead{$10^{22}$\,cm$^{-2}$} &  \colhead{$10^4$ K} & \colhead{ } & \colhead{ } & \colhead{$10^{33}$\,erg\,s$^{-1}$} & \colhead{}   
}
\startdata
P1          & 2006.8 & $2.14 \pm 0.02$ & $200.0^{+0.9}_{-1.0}$ & $7.33 \pm 0.17$  & $2.81 \pm 0.08$  & $6.4 \pm 0.1$ & 1.17 (242)\\ [0.7ex]
P2          & 2012.3 & $=N_H$(P1)      & $199.1 \pm 1.0$       & $7.12 \pm 0.17$  & $2.74 \pm 0.08 $ & $6.3 \pm 0.1$ & 1.17 (242) \\[0.7ex]
P3 (a\,\&\,b)  & 2015.3 & $=N_H$(P1)      & $198.3 \pm 0.8$       & $7.00 \pm 0.13$  & $2.70 \pm 0.07 $ & $6.2 \pm 0.1$ & 1.17 (242) \\[0.7ex]
P4 (a\,\&\,b) & 2020.4 & $=N_H$(P1)      & $196.1 \pm 0.9$       & $6.54^{+0.16}_{-0.17}$  & $2.55^{+0.08}_{-0.07}$ & $5.9 \pm 0.1$ & 1.17 (242) \\[0.7ex]
\hline
P1         & 2006.8 & $2.19 \pm 0.06$ & $200.7 \pm {1.2}$      & $7.39 \pm 0.18$  & $2.88 \pm 0.11 $& $6.5 \pm 0.2$ & 1.13 (239)\\[0.7ex]
P2         & 2012.3 & $2.22 \pm 0.07$ & $200.2^{+1.2}_{-1.3} $ & $7.19 \pm 0.18$  & $2.83\pm 0.12$  & $6.5 \pm 0.2$ & 1.13 (239) \\[0.7ex]
P3 (a\,\&\,b) & 2015.3 & $2.09 \pm 0.06$ & $197.7 \pm 1.1$        & $6.96 \pm 0.14$  & $2.65\pm 0.09 $ & $6.1 \pm 0.1$ & 1.13 (239) \\[0.7ex]
P4 (a\,\&\,b) & 2020.4 & $2.01 \pm 0.09$ & $194.5^{+1.4}_{-1.5}$  & $6.48 \pm 0.17$  & $2.44\pm 0.12 $  & $5.8 \pm 0.2$ & 1.13 (239) \\[0.7ex]
\enddata
\tablecomments{The fits were done simultanously for P1-P4, the parameters are tied for P3a and P3b, and for the two observations of P4. The normalization is fixed for all epochs in all fits at $\mathcal{N}=923$ (see text).
Fluxes are given for the energy range of 0.6-6\,keV. $F^{\rm{abs}}_{-13}$ is the absorbed flux in units of $10^{-13}$\,erg\,cm$^{-2}$\,s$^{-1}$, while $F^{\rm{unabs}}_{-12}$ is the unabsorbed flux in units of $10^{-12}$\,erg\,cm$^{-2}$\,s$^{-1}$. 
All errors indicate the 90\% confidence level for one parameter of interest. 
The bolometric luminosity at inifinity is calculated as $L^{\infty}_{\rm bol}=4 \pi \sigma {R^{\infty}_{\rm Em}}^2 {T^{\infty}_{\rm eff}}^4= 4 \pi \sigma 10^{10} \mathcal{N} d_{\rm 10kpc}^2 {T_{\rm eff}}^4 (1+z)^{-2}$\,erg\,s$^{-1}$. Its uncertainty only considers the uncertainty of the temperature. $t_{\rm Omid}$ indicates the middle of the observation time of the respective epoch. {$\chi^2_{\nu}$  is the reduced $\chi^2$ and dof are the degrees of freedom of the X-ray spectral fits.}\vspace{-0.5cm}}
\end{deluxetable*}

\begin{figure*}[]
{\includegraphics[height=80mm]{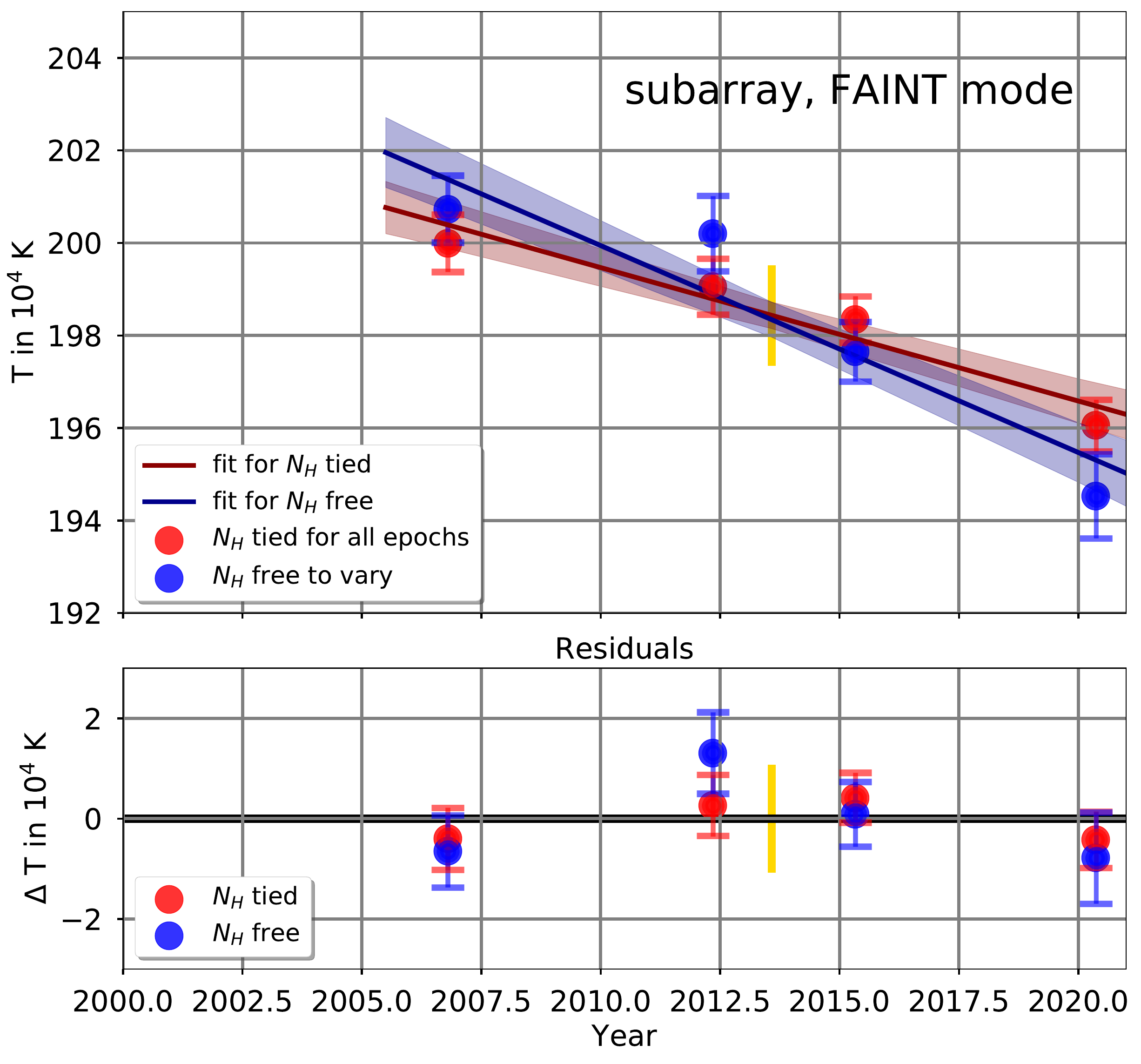}
\includegraphics[height=80mm]{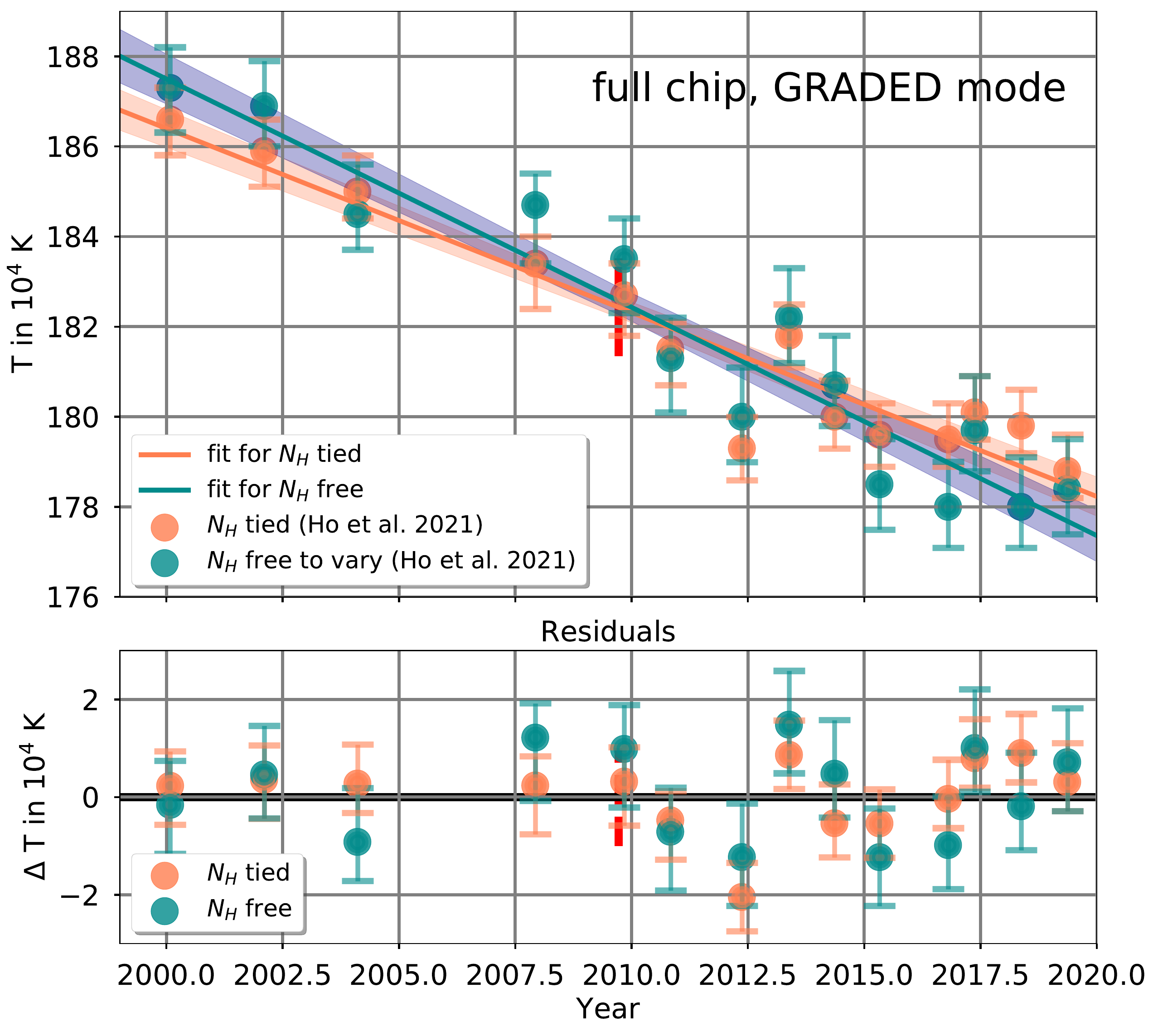} } 
\caption{Temperature change over time (upper panels), and residuals to the linear fits (lower panels). All errors in this plot are $1\sigma$ errors. The different instruments modes resulted in a pileup fraction less than 1.6\% for the analysized data of the left panels \citep{PavlovL2009} and a slowy decreasing pileup fraction from 13\% (2000) to 6\% (2019) according to \citetalias{Ho2021} for the data of the right panels. 
 Our fit results for the subarray data from Table~\ref{table:newcarbonfits} for a carbon atmosphere model are marked with red points (same $N_{\rm H}$ in all epochs) and blue points (different $N_{\rm H}$) in the left panels. 
The respective red (blue) line and shaded area in the left panel indicate the results of a linear regression fit and its $1\sigma$ error to the data points where $N_{\rm H}$ is tied (allowed to vary).
On the right side we show the carbon atmosphere fit results from \citet{Ho2021} for the Graded mode data and the respective linear fit and residuals in coral (cyan) if
$N_{\rm H}$ is tied (allowed to vary). The vertical yellow and red lines mark the chosen reference times, $t_0=2013.58$ and $t^{\rm{H+21}}_0=2009.72$, for the left and right panels respectively.
\label{TevoSG}}
\end{figure*}

As in \citetalias{Posselt2018}, we carried out standard least-square fits to a straight line, $T_{\rm eff}= T_0 + \dot{T}(t-t_0)$, where $t$ is the time of observation and  $t_0=2013.58$ the chosen reference time, in the middle of all subarray observations. 
If $N_{\rm H}$ is allowed to vary between epochs, we derive a slope
$\dot{T}=-4500 \pm 800$\,K\,yr$^{-1}$ and an intercept $T_0=(198.3 \pm 0.4)\times 10^4$\,K ($1\sigma$ uncertainties, $\chi^2_{\nu}=2.1$ for $\nu=2$\,dof), shown by the blue points and area in Figure~\ref{TevoSG}. This corresponds to a cooling rate of $-2.3 \pm 0.4 (1\sigma)$\,\% in 10 years, and to a characteristic cooling time, $\tau_{\rm cool}= T_0/(-\dot{T})=440\pm 80$\,yr.
If $N_{\rm H}$ is the same for all epochs, the values are  $\dot{T}=-2900\pm 600$\,K\,yr$^{-1}$, $T_0=(198.4 \pm 0.3)\times 10^4$\,K ($1\sigma$ uncertainties, $\chi^2_{\nu}=0.9$ for $\nu=2$\,dof), shown by the red points and area in Figure~\ref{TevoSG}. This corresponds to a cooling rate of $-1.5 \pm 0.3 (1\sigma)$\,\% in 10 years, and to a characteristic cooling time, $\tau_{\rm cool}= 690 \pm 140$\,yr.\\ 

{The fit for $T(t)$ is worse for the case} where
$N_{\rm H}$ is allowed to vary between epochs, and stronger residuals are apparent in Figure~\ref{TevoSG} (left panels).
This is due to the seemingly faster temperature change between the two last epochs 
in comparison to the first two epochs as already mentioned above and visible in Figure~\ref{carbonNHT}. 
This can be also illustrated by only considering the last 3 epochs for a linear temperature fit. For such a fit, the best-fit temperature value in epoch 1 (Table~\ref{table:newcarbonfits}) is $4.3\sigma$
 away from the temperature one would expect according to the linear fit parameters ($\dot{T}_{234}=-7000 \pm 1500$\,K\,yr$^{-1}$, $T_{0,234}=(197.2 \pm 0.5)\times 10^4$\,K; $1\sigma$ uncertainties, $t_{0,234}=2016.4$).
If the fit results for the varying $N_{\rm H}$ are taken at face value, it means that either  epoch 1 (2006) is an outlier or the cooling of the CCO  accelerated after 2012. 
We regard the latter scenario as unlikely.
In 2006, however, the optical thickness of the contamination on the ACIS blocking filter was still low and the instrument sensitivity {was} the best of the four epochs. Thus, a deviation of epoch 1 would be also puzzling. The fit where $N_{\rm H}$ is tied for all epochs is statistically acceptable. For both, $N_{\rm H}$ free or tied, we note that most of the temperature drop comes from the last epoch - as Figure~\ref{carbonNHT} illustrates, in the last 5 years the differences are as large (or slightly larger) than the respective ones over the first 9 years. 
If only the first three epochs with the new CALDB are considered, we obtain $\dot{T}_{123}=-1900\pm 900 (1\sigma)$\,K\,yr$^{-1}$ (tied $N_H$), and $\dot{T}_{123}=-3400\pm 1100(1\sigma)$\,K\,yr$^{-1}$ (free $N_H$), i.e., lower than the values for all 4 epochs above.\\

Our results are within the $\dot{T}$-bounds reported by \citetalias{Posselt2018}. 
The ten-year cooling rates are consistent with (although slightly slower than) the respective values recently presented by \citetalias{Ho2021} ($-2.2 \pm 0.2$ or $-2.8 \pm 0.3$\,\% per 10\,yr, corresponding to $\dot{T}^{\rm{H+21}}=-4090 \pm 360$\,K\,yr$^{-1}$ and $\dot{T}^{\rm{H+21}}=-5070 \pm 480$\,K\,yr$^{-1}$  for tied or varying $N_H$, respectively; {all with their respective $1\sigma$ uncertainties}) based on 19\,years of Graded mode data of the Cas A supernova remnant.   
Figure~\ref{TevoSG} shows the results on the temperature change from our study (left panels) and the \citetalias{Ho2021} study (right panels) together. 
Only the relative changes are relevant. The offsets in absolute temperatures are due to different normalizations (reflecting different radius and mass assumptions, spectral model normalization, scattering and pileup considerations for \citetalias{Ho2021}).
Interestingly, the last four \citetalias{Ho2021} epochs seem to indicate a slowing of the temperature decrease -- the opposite behavior to what our best-fit values seem to imply. 
In addition, the \citetalias{Ho2021} residuals closest in time to the time of our residuals show nearly mirrored behavior. For instance, our residuals in 2012 are positive, while the \citetalias{Ho2021} residuals of 2012 are negative.
Although a bit surprising, not much can be learned from this since statistical fluctuations can explain both of these (insignificant) trends.\\    

As a final note, we emphasize that the non-magnetic carbon atmosphere model is not the only one that fits the CCO spectrum. For instance, hydrogen atmosphere models with  low magnetic fields ($B < 10^{10}$\,G) fit the CCO spectra equally well. Such a fit does not show a temperature decrease over time \citepalias{Posselt2013}. The fit with the hydrogen atmosphere models implies small emission areas, i.e. hot spot emission{, and the apparent flux decrease is due to decreasing hot spot area.} In contrast, a fit with a carbon atmosphere model produces an emission size consistent with expectations for the entire surface of a neutron star.  
Since no X-ray pulsations of the CCO (as one might expect for hot spots) have been detected, and the hypothesis of residual nuclear burning for very young and hot neutron stars can be applied to this CCO, it is argued  that a carbon atmosphere appears likely for the Cas\,A CCO (see, e.g., \citealt{Wijngaarden2019}).
We note that the used carbon atmosphere models are all non-magnetic ($B<10^8$\,G), and even a relatively weak field
strength ($B < 10^{10}$\,G) can affect the emergent model spectrum \citep{Suleimanov2014}.
Unfortunately, the currently available X-ray data do not allow one to directly differentiate between hydrogen and carbon atmosphere models, so a hydrogen atmosphere remains a possibility.

\section{Summary}
We used four epochs of \emph{Chandra} observations of the Cas A CCO in the ACIS subarray mode during a time span of 14 years to constrain the change of the temperature as $\dot{T}=-2900\pm 600$\,K\,yr$^{-1}$  ($\dot{T}=-4500\pm 800$\,K\,yr$^{-1}$) corresponding to cooling rates of $-1.5 \pm 0.3$\,\% per 10\,yr ($-2.3 \pm 0.4$\,\% per 10\,yr) for the assumption of the same (or varying) $N_H$. The fit statistics and the residuals of the linear $T(t)$ regression indicate a better fit in the case of the same $N_H$. At the same time, if $N_H$ is allowed to vary, its values can deviate by more than $3\sigma$, calling into question the assumption of the same $N_H$ value for all epochs.\\

The best-fit temperature changes are obtained under the following assumptions: (i) the Cas A CCO has a {non-magnetic} carbon atmosphere that covers the entire neutron star surface, (ii) the effective temperature is uniformly distributed over the surface, i.e., there are no hot spots,
(iii) the calibration data base CALDB 4.9.5 fully corrects for all instrument effects, in particular the time-dependent effects of the accumulating contamination layer on the ACIS optical blocking filter.\\
 
According to our spectral fit results, the temperature decrease in the last 5 years is as large (or even slightly larger) than the temperature drop over the first 9 years. Together with the findings for the varying $N_H$, we caution that assumption (iii) may not be correct and further improvements to the time-dependent \emph{Chandra} calibration may be needed.\\ 

Our results on the temperature slope are consistent with the results obtained by \citetalias{Ho2021} from \emph{Chandra} ACIS Graded mode observations. Of course, their data are subject to the same assumptions. In addition, the necessary correction for the pileup effect of the full-chip data implies an additional assumption regarding the instrument effects. While our data show the largest apparent temperature drop over from 2015 to 2020, the \citetalias{Ho2021} data show a surprisingly small temperature change during the same time. These different behaviors in the general temperature evolution can still be attributed to statistical fluctuations. However, they also indicate that the current status of the data and calibration should not be the end of the quest to constrain the temperature evolution of the youngest known CCO.  

\begin{acknowledgments}
The scientific results reported in this article are based on observations made by the \emph{Chandra} X-ray Observatory.
Support for this work was provided by the National Aeronautics and Space Administration through Chandra Award Number GO0-21049X issued by the Chandra X-ray Observatory Center, which is operated by the Smithsonian Astrophysical Observatory for and on behalf of the National Aeronautics Space Administration under contract NAS8-03060. 
BP acknowledges funding from the UK Science and Technology Facilities Council (STFC) Grant Code ST/R505006/1.

\end{acknowledgments}

\bibliographystyle{aasjournal}
\bibliography{Casa}

\end{document}